\documentstyle[amssymb,aps,multicol,graphicx]{revtex}

\begin{document}

\title{The Excess Wing in the Dielectric Loss of Glass-Formers: A Johari-Goldstein $\beta $-Relaxation?}
\author{U. Schneider, R. Brand, P. Lunkenheimer, and A. Loidl}
\address{Experimentalphysik V, Universit\"{a}t Augsburg, D-86135 Augsburg, Germany}
\date{submitted to Phys.~Rev.~Lett. 05.01.2000; revised 02.03.2000}
\maketitle

\begin{abstract}
Dielectric loss spectra of glass-forming propylene carbonate and glycerol at temperatures above and below $T_{g}$ are
presented. By performing aging experiments lasting up to five weeks, equilibrium spectra below $T_{g}$ have been
obtained. During aging, the excess wing, showing up as a second power law at high frequencies, develops into a
shoulder. The results strongly suggest that the excess wing, observed in a variety of glass formers, is the
high-frequency flank of a ${\rm \beta }$-relaxation.
\end{abstract}


\begin{multicols}{2}
\columnseprule0pt \narrowtext

In the frequency-dependent dielectric loss, $\varepsilon "(\nu )$, of many glass-formers, an excess contribution to
the high-frequency power law of the ${\rm \alpha }$-peak, $\varepsilon "\sim \nu ^{-\beta }$, shows up about 2-3
decades above the ${\rm \alpha }$-peak frequency $\nu _{p}$. This excess wing can be reasonably well described by a
second power law, $\varepsilon "\sim \nu ^{-b}$, with $b<\beta $ \cite{Ho94,Lu96a,Le97a}. An excess contribution at
high frequencies was already noted in the pioneering work of Davidson and Cole \cite{CD} and it is a common feature in
glass-forming liquids without a well-resolved ${\rm \beta }$-relaxation (see below). The physical origin of the excess
wing is commonly considered as one of the great mysteries of glass physics. Despite the fact that some theoretical
approaches describing the excess wing have been proposed \cite{wingtheo}, up to now there is no commonly accepted
explanation of its microscopic origin. In addition, some phenomenological descriptions have been proposed \cite
{Le97a,Di90a,Ku99}. Most successful is the so-called Nagel scaling \cite {Di90a}, which leads to a collaps of the
$\varepsilon "(\nu )$ curves (including the excess wing) for different temperatures and various materials onto one
master curve.

In various glass-formers additional relaxation processes, usually termed ${\rm \beta }$-processes, are clearly seen in
the loss spectra, showing up as a shoulder or even a second peak at $\nu
>\nu _{p}$ \cite{Ku99,Jo70}. Such relaxations in supercooled liquids without intramolecular degrees of freedom were
systematically investigated already three decades ago by Johari and Goldstein \cite{Jo70} and basically were ascribed
to intrinsic relaxation phenomena of the glass-forming process. We recall that these Johari-Goldstein ${\rm \beta
}$-relaxations are fundamentally different from ${\rm \beta }$-relaxations due to intramolecular degrees of freedom.
Commonly it is assumed that the excess wing and ${\rm \beta }$-relaxations are different phenomena \cite{Di90a,Ku99}
and even the existence of two classes of glass-formers was proposed - ''type A'' with an excess wing and ''type B''
with a ${\rm \beta }$-process \cite{Ku99}. Based on experimental observations it seems natural to explain
Johari-Goldstein ${\rm \beta }$-processes and excess wing phenomena on the same footing. Indeed such a possibility was
already considered earlier \cite {Ho94,Olsen,Jimenez99,Lerapidity,Le110,Wagner99} and from an experimental point of
view, it cannot be excluded that the excess wing is simply the high-frequency flank of a ${\rm \beta }$-peak, hidden
under the dominating ${\rm \alpha }$-peak.

However, only the presence of a shoulder with a downward curvature can provide an unequivocal proof for an underlying
${\rm \beta }$-peak. In materials with a well pronounced ${\rm \beta }$-peak, the ${\rm \alpha }$- and ${\rm \beta
}$-relaxation timescales increasingly separate with decreasing temperature. Therefore, within this picture it may be
expected that at low temperatures the excess wing transfers into a shoulder, due to less interference with the ${\rm
\alpha }$-peak. In marked contrast to this prediction, for various glass-forming liquids far below the glass
temperature $T_{g}$, $\varepsilon "(\nu )$ follows a well-developed power law, without any indication of a shoulder
\cite{Ho94,Ku99}. However, measurements of glass-forming materials at low temperatures near or below $T_{g}$ suffer
from the fact that thermodynamic equilibrium is not reached in measurements using usual cooling rates. In addition,
the loss becomes very low yielding increasingly high experimental uncertainties. Clearly, high-precision equilibrium
measurements are necessary to investigate the ''true'' behavior of the excess wing at low temperatures.

Therefore we performed high-precision dielectric measurements keeping the sample at a constant temperature, some ${\rm
K}$ below $T_{g}$, for up to five weeks which ensured that thermodynamic equilibrium was indeed reached. The
experiments were performed on two prototypical glass-formers known to exhibit well pronounced excess wings, propylene
carbonate (PC, $T_{g}\approx 159~{\rm K}$) and glycerol ($T_{g}\approx 185~{\rm K}$). For the measurements, parallel
plane capacitors having an empty capacitance up to $100~{\rm pF}$ were used. Due to the approach of the fictive
\cite{fictive} towards the real temperature during aging, a thermal contraction of the sample may occur. Its magnitude
in one dimension can be estimated to $\Delta l/l\leq 3\times 10^{-3}$,~which has a negligible effect on the results.\
High-precision measurements of the dielectric permittivity in the frequency range $10^{-4}~{\rm Hz}\leq \nu \leq
10^{6}~{\rm Hz}$ were performed using a Novocontrol alpha-analyser. It allows for the detection of values of $\tan
\delta $ as low as $10^{-4}$. At selected temperatures and aging times, additional frequency sweeps at $20~{\rm
Hz}\leq \nu \leq 1~{\rm MHz}$ were performed with the autobalance bridges Hewlett-Packard HP4284 and HP4285. To keep
the samples at a fixed temperature for up to five weeks, a closed-cycle refrigerator system was used. The

\begin{figure}[tbp]
 \includegraphics[clip,width=8cm]{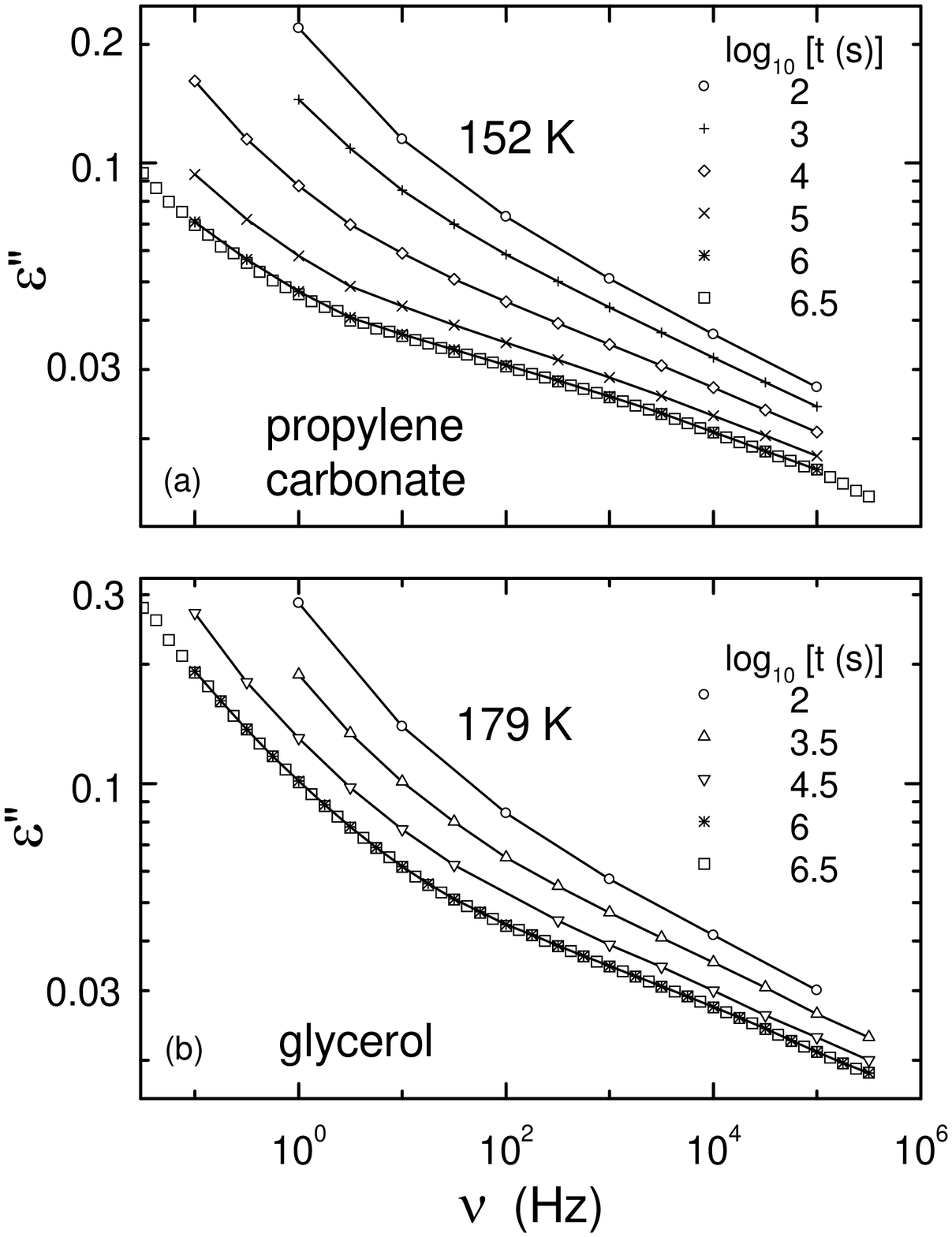}
 \caption{Frequency-dependent dielectric loss of PC (a) and glycerol (b) for different aging times $t$ (not
all shown). The lines connect the symbols for $t\leq 10^{6}~{\rm s}$. The spectra for the maximum aging time of
$10^{6.5}~{\rm s}$ ($\square $) have been taken in an extended frequency range.} \label{fig1}
\end{figure}

\noindent sample was cooled from a temperature $20~%
{\rm K}$ above $T_{g}$\ with the maximum possible cooling rate of about $3~%
{\rm K/min}$. The final temperature was reached without any temperature undershoot. As zero point of the aging times
reported below, we took the time when the desired temperature was reached, about $200~{\rm s}$ after passing $T_{g}$.
The temperature was kept stable within $0.02~{\rm K}$ during the five weeks measurement time.

Figure 1 shows $\varepsilon "(\nu )$ of PC and glycerol at temperatures some ${\rm K}$\ below $T_{g}$ during aging. A
strong dependence of the spectra on aging time $t$ is observed. The spectra measured after the maximum time of
$t=10^{6.5}~{\rm s}$ (squares) are identical to those obtained after $t=10^{6}~{\rm s}$ (stars), clearly demonstrating
that equilibrium was reached. In Fig. 2 these equilibrium spectra are shown, together with equilibrium curves taken at
higher temperatures, partly published earlier \cite{Sc}. Obviously, in the frequency window of Fig. 1 mainly the excess
wing region is covered; the ${\rm \alpha }$-peak is located at much lower frequencies, leading to a somewhat steeper
increase towards low frequencies only. In Fig. 1 for both materials indeed the typical power law, characteristic of an
excess wing, is observed for short $t$. However, with increasing $t$ the excess wing successively develops into a
shoulder! This behaviour strongly suggests that a relaxation is the origin of the excess wing observed at shorter
times or

\begin{figure}[tbp]
 \includegraphics[clip,width=8cm]{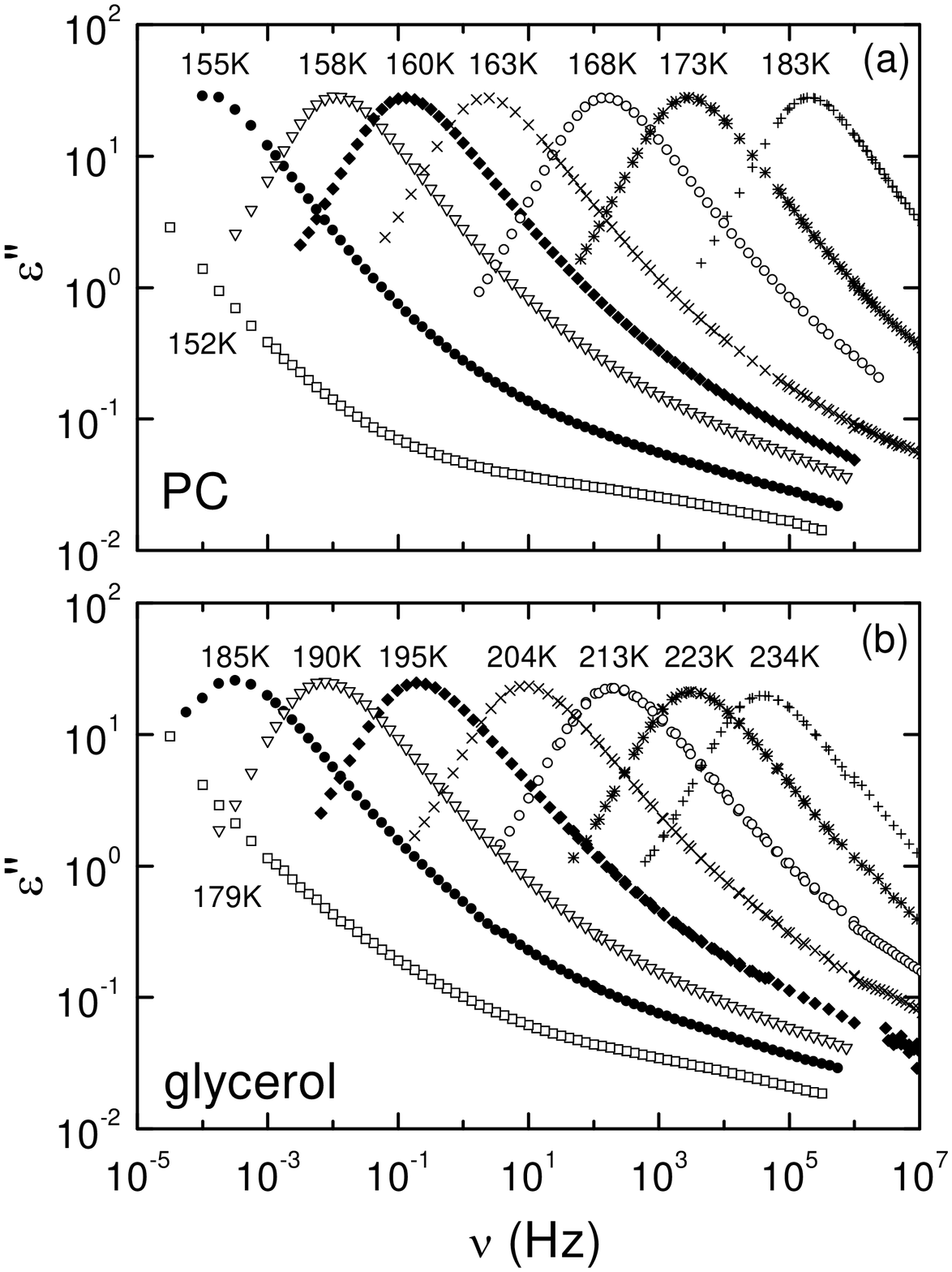}
 \caption{Frequency-dependent dielectric loss of PC (a) and glycerol (b) for different temperatures. All curves were
taken in thermodynamic equilibrium. } \label{fig2}
\end{figure}

\noindent higher temperatures (Fig. 2). Admittedly, for glycerol the curvature in $\log _{10}\varepsilon "(\log
_{10}\nu )$ is quite subtle (Fig. 1b). However, further evidence is provided by Fig. 3 where the derivatives of the
lowest curves in Figs. 1 and 2 are shown. A maximum is observed, clearly indicative of a ${\rm \beta }$-process \cite
{Ku99}. In the region of the shoulder, the values of $\tan \delta $\ ranging between $5\times 10^{-3}$ and $2\times
10^{-2}$ are well above the resolution limits of the measurement. In addition, for both materials measurements using
the autobalance bridges (not shown) agree well with those obtained with the alpha-analyser. Finally, the response of
the empty capacitor was measured at the temperatures of Fig. 1. The resulting loss of $\varepsilon "<1.5\times
10^{-4}$ in the region of $10~{\rm Hz}\leq \nu \leq 100~{\rm kHz}$ excludes a non-intrinsic origin of the observed
shoulder.

The results of Fig. 1\ are interpreted as follows: During aging the fictive temperature \cite{fictive} successively
approaches the ''real'' temperature leading to a strong shift of the ${\rm \alpha }$-peak towards lower frequencies.
The relaxation times of ${\rm \beta }$-processes usually exhibit a weaker temperature dependence than those of the
${\rm \alpha }$-process. Therefore the ${\rm \beta }$-process causing the excess wing can be expected to be less
affected by the aging. The different aging depencences of the ${\rm \alpha }$- and the ${\rm \beta }$-process can also
easily be deduced from the fact that the spectra at different aging times cannot be scaled onto each other by a
horizontal shift in Fig. 1. In
addition, aging experiments reported by Johari and

\begin{figure}[tbp]
 \includegraphics[clip,width=7cm]{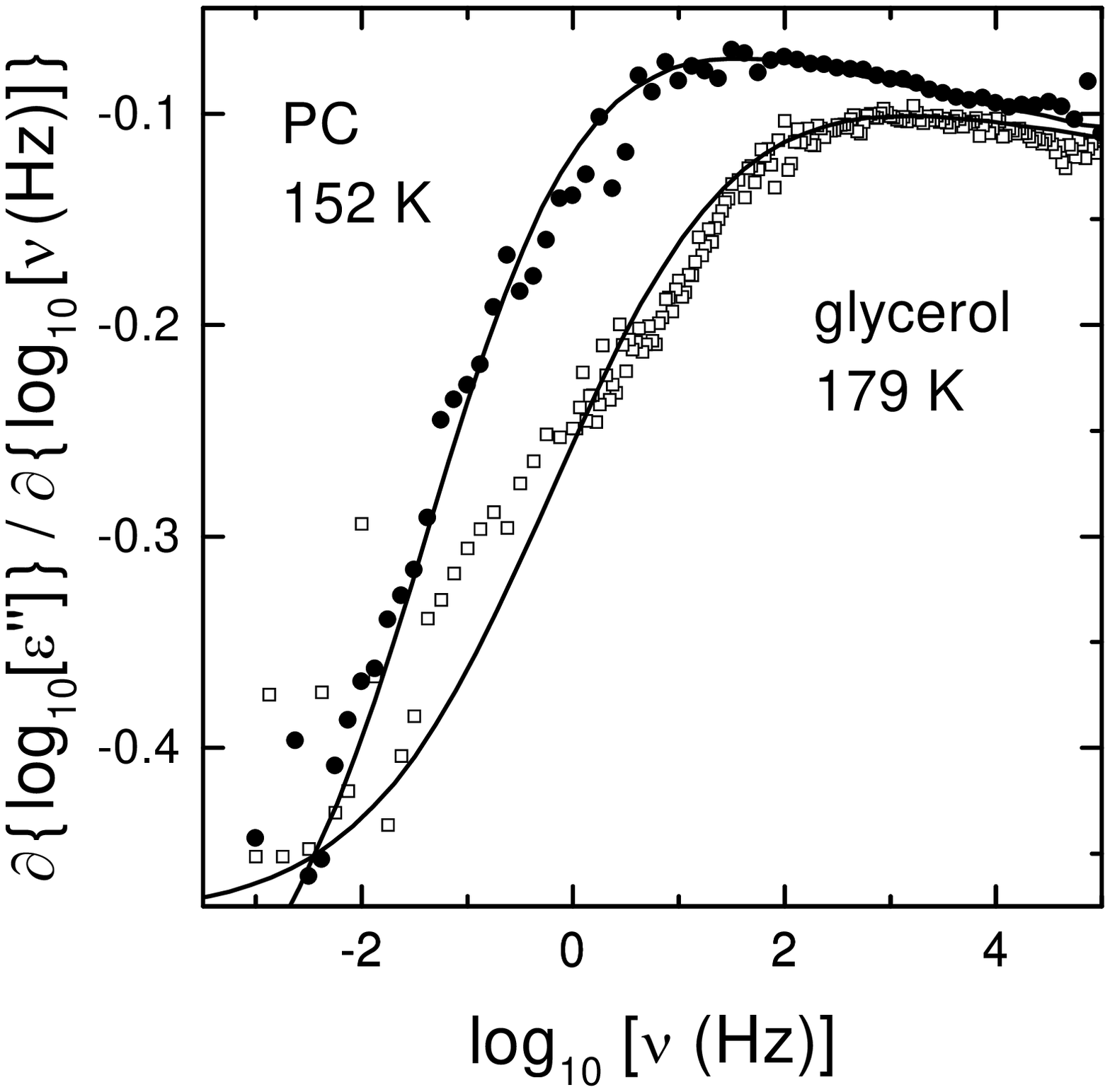}
 \caption{Derivative of the dielectric loss spectra obtained after $10^{6.5}~{\rm s}$. The lines correspond to the fits
shown in Fig. 4. } \label{fig3}
\end{figure}

\noindent coworkers \cite{Jo70} revealed a negligible change of the ${\rm \beta }$-peak frequency with aging time
\cite{remolsen}. Overall, with increasing aging time, respectively decreasing effective temperature, the time scales
of ${\rm \alpha }$- and ${\rm \beta }$-process become successively separated and finally the presence of a ${\rm \beta
}$-relaxation is revealed by the appearence of a shoulder.

A close inspection of dielectric measurements at low temperatures,\ reported in literature, reveals some further
evidence of a ${\rm \beta }$-relaxation being the origin of the excess wing: Already in the spectrum obtained by
Leheny and Nagel \cite{Leheny} after aging of glycerol for $2.5\times 10^{6}~{\rm s}$ at $177.6~{\rm K}$ a vague
indication for a shoulder is present. However, possibly due to doubt of its significance, these authors payed no
attention to this feature. In addition, in various other published results, near or just below $T_{g}$\ some
indications for a shoulder show up \cite {Ho94,Le97a,Wagner99,Sc,Ku95}, again being ignored in most of these
publications. Except for \cite{Wagner99} the thermal history was not reported in detail and it can be doubted if
thermal equilibrium was reached. In the study of Wagner and Richert \cite{Wagner99}, in highly quenched Salol a strong
secondary peak appeared below $T_{g}$ which vanished after heating to $T_{g}$. In addition, in isochronal measurements
of various quenched glass-formers \cite{Hansen97} non-equilibrium secondary relaxations were detected. From the
results in \cite{Wagner99,Hansen97} the question arises if the subtle indications for a ${\rm \beta }$-relaxation seen
around $T_{g}$ in \cite{Ho94,Le97a,Wagner99,Sc,Ku95} are pure non-equilibrium effects and if this relaxation may
completely vanish in equilibrium. Only the present aging experiments could prove that the excess wing is due to a
${\rm \beta }$-relaxation that is an equilibrium phenomenon. In Fig. 1 the amplitude of the ${\rm \beta }$-peak (only
seen as excess wing at short times) may be suspected to diminish with aging. Interestingly, in systems with well
resolved ${\rm \beta }$-relaxations a similar behavior was observed \cite {Jo70,Olsen}. Finally it should be mentioned
that in \cite{Jo86} evidence of a ${\rm \beta }$-relaxation in various systems without a clearly resolvable second
relaxation peak (including PC) was found by using a ''difference isochrone'' method.

\begin{figure}[tbp]
 \includegraphics[clip,width=7cm]{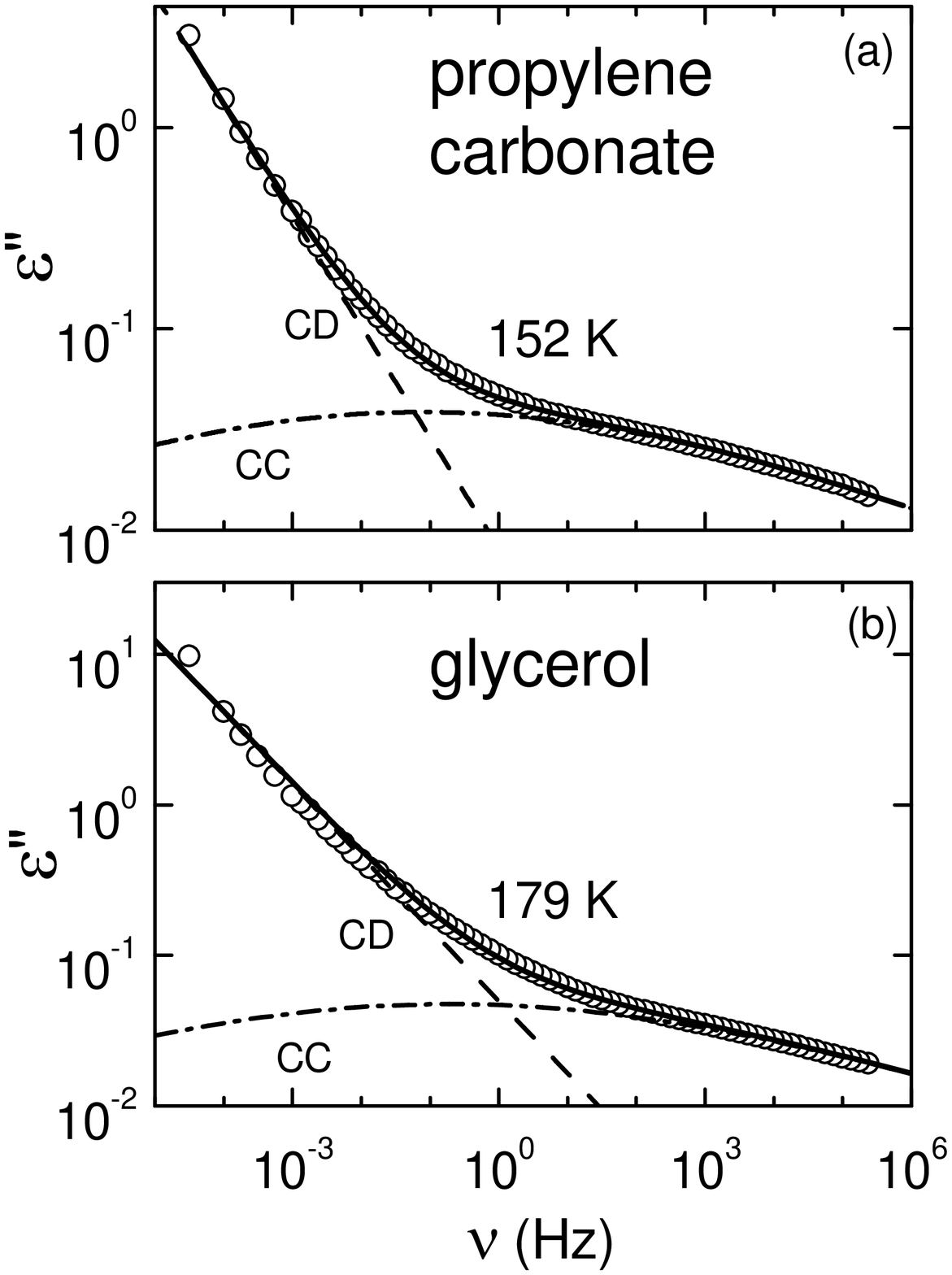}
 \caption{Dielectric loss spectra obtained after $10^{6.5}~{\rm s}$. The\ solid lines are fits with a sum of a CD
(dashed line) and a CC function (dash-dotted).} \label{fig4}
\end{figure}

\noindent In addition, in glycerol a high-frequency shoulder was detected by performing high-pressure dielectric
experiments \cite {Joharipress}. Further evidences for ${\rm \beta }$-relaxations in typical excess-wing glass-formers
have also been obtained from other experimental methods, e.g. calorimetry \cite{Fujimori95} or nuclear magnetic
resonance \cite{Sc92}. A strong corroboration of the present results is also provided by our recent results in
ethanol, which prove that the well-pronounced excess wing in this glass-former is also due to a ${\rm \beta
}$-relaxation \cite{eth}.

What further information on the ${\rm \beta }$-relaxation causing the excess wing can be gained from the present data?
The ${\rm \alpha }$-peaks in PC and glycerol can well be fitted using the empirical Cole-Davidson (CD) function
\cite{CD,Sc}. In materials with well pronounced ${\rm \beta }$-peaks their spectral shape often can be described by
the Cole-Cole (CC) function \cite{Co41}. In Fig. 4 we demonstrate that the equilibrium curves of Fig. 1 can well be
fitted using a sum of a CD and a CC function. Similar analyses were performed in
\cite{Ho94,Jimenez99,Lerapidity,Le110}. In Fig. 3 the corresponding derivatives are shown -- again a reasonable
agreement can be stated. In the same way also the present spectra at higher temperatures (Fig. 2) can be described
\cite{LunkiHab}. The resulting characteristics of the ${\rm \beta }$-relaxations in PC and glycerol can be summarized
as follows: i) The relaxation strength increases with $T$. This is often found for well resolved ${\rm \beta
}$-relaxations, e.g. in \cite{Ku99}. ii) The width of the relaxation peaks becomes smaller with $T$, which is also a
common property of well separated ${\rm \beta }$-relaxations \cite{Ku99} iii) Despite some difficulty to obtain
precise information on the temperature development of the ${\rm \beta }$-relaxation time, $\tau _{\beta }$, the
analysis reveals significant deviations from thermally activated behavior \cite{LunkiHab}. Therefore one may have
objections using the term ''${\rm \beta }$-relaxation'' in this case, as they are commonly assumed to follow an
Arrhenius behavior. However, there is no principle reason that ${\rm \beta }$-processes always should behave thermally
activated, especially as their microscopic origin is still unclear. Already Johari \cite{Jo70} suspected that in
systems without a well resolved ${\rm \beta }$-process, the relaxation times of ${\rm \alpha }$- and ${\rm \beta
}$-process are closer together due to an uncommon temperature dependence of $\tau _{\beta }$. Maybe the only
difference between ''type A'' and ''type B'' glass-formers in the classification scheme of \cite{Ku99} is just the
temperature evolution of the ${\rm \beta }$-dynamics: In the latter materials it is thermally activated, leading to a
clear separation of ${\rm \alpha }$- and ${\rm \beta }$-peak at low temperatures. In contrast, in ''type A'' materials
$\tau _{\beta }(T)$ follows more closely $\tau _{\alpha }(T)$ (i.e., it deviates from Arrhenius) and only the
high-frequency flank of the ${\rm \beta }$-peak -- the excess wing -- is visible. The Nagel-scaling \cite {Di90a},
which strongly suggests a correlation between the ${\rm \alpha }$-process and the excess wing in ''type A'' systems,
is not opposed to this framework: It simply implies that in those materials the parameters characterizing both
relaxation processes are intimately related.

In this context it is of interest that recently for ''type B'' glass-formers
a correlation of $\tau _{\beta }$ and the Kohlrausch-exponent $\beta _{KWW}$
(describing the width of the ${\rm \alpha }$-peak), both at $T_{g}$, was
found \cite{Ngaibeta}: log$_{10}\tau _{\beta }(T_{g})$ increases
monotonically with $\beta _{KWW}(T_{g})$. It was noted that those
glass-formers that show no well-resolved ${\rm \beta }$-relaxation
(including PC and glycerol) have relatively large values of $\beta
_{KWW}(T_{g})$. As the mentioned correlation implies that the ${\rm \alpha }$%
- and ${\rm \beta }$-timescales approach each other with increasing $\beta
_{KWW}(T_{g})$, the unobservability of a shoulder in those materials is
easily rationalized. In \cite{Ngaibeta} an explanation of this behavior
within the coupling model (CM) \cite{CM} was proposed. Recently it was shown
that the Nagel-scaling can be explained within this framework \cite
{Lerapidity} and even the deviation of $\tau _{\beta }(T)$ from thermally
activated behavior seems to be consistent with the CM.

In summary, by performing high-precision measurements below $T_{g}$ in
thermodynamic equilibrium we were able to resolve the presence of ${\rm %
\beta }$-relaxations in two typical excess wing glass-formers. Our results
strongly suggest that the so-far mysterious excess wing, showing up at
higher temperatures, is simply the high-frequency flank of the loss-peaks
caused by these ${\rm \beta }$-relaxations. Of course, the mere
identification of the excess wing with a ${\rm \beta }$-relaxation provides
no explanation, e.g. for\ the Nagel scaling \cite{Di90a} or the question why
glass-formers seem to fall into two classes - those with a well-separated $%
{\rm \beta }$-relaxation and those without \cite{Ku99}. But the knowledge of
the true nature of the excess wing certainly is a prerequisite for any
microscopic explanation of these phenomena, (e.g. \cite{Lerapidity,Ngaibeta}%
) and hopefully will enhance our understanding of the glass transition in
general.

We thank K.L.~Ngai for stimulating dicussions. This work was supported by
the DFG, Grant-No. LO264/8-1 and partly by the BMBF, contract-No. EKM
13N6917.

\end{multicols}


\begin{references}
\bibitem{remolsen}  In \cite{Olsen} even a small upward shift of the ${\rm %
\beta }$-peak frequency was noted during aging.
\end{references}
\end{document}